\documentclass[journal]{IEEEtran}
\usepackage{graphicx}
\usepackage{amsmath}
\ifCLASSINFOpdf

\else

\fi

\begin{document}

\title{A decentralized approach towards secure firmware updates and testing over commercial IoT Devices}

\author{Projjal~Gupta,~\IEEEmembership{Member,~IEEE,}

\thanks{P. Gupta is with the Department
of Electronics and Communications Engineering, SRM Institute of Science and Technology, Kattankulathur, Tamil Nadu, India e-mail: projjalgupta@hotmail.com.}
}

\maketitle

\begin{abstract}
Internet technologies have made a paradigm shift in the fields of computing and data science and one such paradigm defining change is the ‘Internet of Things’ or IoT. Nowadays, thousands of household appliances use integrated smart devices which allow remote monitoring and control and also allow intensive computational work such as high end AI-integrated smart security systems with sustained alerts for the user.  The update process of these IoT devices usually lack the ability of checking the security of centralized servers, which may be compromised and host malicious firmware files as it is presumed that the servers are secure during deployment. The solution for this problem can be solved using a decentralized database to hold the hashes and the firmware.
This paper discusses the possible implications of insecure servers used to host the fimrwares of commercial IoT products, and aims to provide a blockchain based decentralized solution to host firmware files with the property of immutability, and controlled access to the firmware upload functions so as to stop unauthorized use. The paper sheds light over possible hardware implementations and the use of cryptographically secure components in such secure architecture models.
\end{abstract}


%
\IEEEpeerreviewmaketitle

\section{Introduction}
Many households have started using smart devices in their daily lives to automate appliances, use personal assistants and other smart facilities such as predictive energy saving, automated garden maintenance etc. This means that all these devices tend to gather a lot of user data which includes personal preferences, account details and family outgoing/incoming log data. This data, in malicious hands, can have devastating effects for the user. Any infiltrator is able to modify the firmware of a given device, and unsafely update it to gain access to all web hooks and can easily gain other personal account details, thereby bypassing all possible software and hardware safeguards by spoofing the hardware security checks. These forms of attacks need to be stopped for the safety and security of its users.
IoT Devices support OTA (Over-the-air) updates in their life cycle, to regularly fix bugs and apply patches to the existing firmware. However, security of the device is flawed if the server side firmware database is compromised. As secure boot and pre-flash checks can be easily emulated and masked by using new hash values (saved in the compromised server), it is unreliable and can be misused. The following paper aims to design a decentralized framework to securely store the firmware files, which is immutable and timestamped. The project aims to add a crypto co-processor to the embedded system solution to allow secure transactions and allows the random check of boot-partitions during secure boot procedure to maintain integrity.

\section{Preface to Security and OTA Services}
\subsection{Possible Attacks and Vulnerabilities}
With the increasing online presence of user data and files, security and privacy of these entities are being taken quite seriously. Due to the existence of private data, there is an ever-increasing fear of a data breach. Nowadays, websites hosted by big companies tend to rely on a semi-centralized network architecture, having multiple copies of a central server, and a load balancer to direct traffic evenly over these set of servers to manage traffic. Many powerful fail safes are employed against denial-of-service (DOS) attacks and Distributed-DOS attacks. As security is of paramount importance, the safeguards are quite potent. Malicious users may try to attack these servers, however, it is still a tough task.
While keeping the security of such servers in mind, many downloadable files are hosted in external servers, distributed over multiple spaced-out geographical locations, to decrease latency problems that may occur for the users accessing them. Similarly, content delivery network (CDN) systems are used for storing data, and these provide an end-point for the users to access it. Most embedded systems and smart devices with over-the-air update (OTA) services use these CDNs to increase the distribution throughput of any new update being pushed to the smart devices.  
In the possible event of a security lapse in these external CDN servers, or the external host, the dearth of security safeguards can lead to a huge set of malicious activities. If a hacker is able to control the distribution of updates, then any new patch can be stopped from being applied to a compromised/vulnerable device. Similarly, if the hacker introduces a new patch, which is a spoof update which can extract personal preferences and details from the local storage of the smart devices, then this can have disastrous consequences. Mainly these can be manipulated by:

\subsubsection{Port Access control spoofing}
Spoofing the port numbers to not allow request to be sent or received
\subsubsection{Denial of Service (DOS)}
Completely blocking any requests and responses to the device
\subsubsection{Man in the middle (MITM) Attack}
Read the data passing to the servers. The data could  be harmless, but many a times the data contains access keys and/or private data and function Identifiers.
\subsubsection{Request/Response Forging}
Forge the requests and responses to completely control the device. This gives the hacker full access to a huge set of devices which can communicate with said server.

\subsection{Analysis of OTA Service}
Over-the-Air (OTA) update is a mechanism employed by interent connected devices or hardware to update settings or device firmware, solely through internet, and not by hardware programming. It a core part of system architecture, with the hardware being able to apply updates correctly, and a server to facilitate this process. 
For IoT devices, multiple robust OTA update methods have been implemented. This is quite important concerning that traditional programming methods cannot be used in the field as regularly and cannot be scaled further in the field.


\begin{figure}[h!]
  \includegraphics[width=\linewidth]{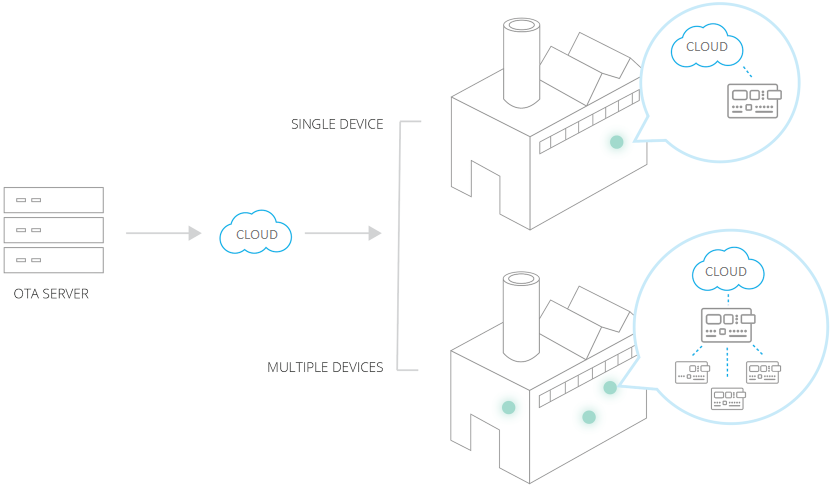}
  \caption{Cloud to Edge/Gateway Architecture in OTA Service for IoT Nodes}
  \label{fig:PIC1}
\end{figure}

\subsection{OTA Architecture for IoT Devices}
IoT devices are designed on the basis of use-case, durability, deployment constraints and various other properties, due to which there can't be any one OTA architecture agnostic for all devices. There are namely three type of OTA architectures:
\subsubsection{Edge-Gateway-Cloud OTA updates}
This architecture uses an internet connected gateway to drive a fleet of smart devices, which in turn is connected to a single cloud storage platform to push updates.

\subsubsection{Gateway-Cloud OTA Updates}
Cloud directly controls the update procedure and the gateway plays no active role in deciding/controlling the update procedures. The entire fleet is partially or fully updated based on commands sent by the cloud.

\subsubsection{Edge-Cloud OTA Updates}
The edge devices (IoT device itself) directly connects to the cloud instance which stores the new firmware. It automatically checks for updates and patches from the cloud, and can also support direct firmware push service from the cloud.

\subsection{OTA Design Considerations for IoT Systems }
When designing OTA updation systems for any embedded systems application, we need to have some key considerations, which will make sure that the procedure is is fail-proof and tolerant towards corrupted firmware installs. The main considerations are:

\subsubsection{Automatic recovery from corrupted or Interrupted updates is a must}
OTA updates should completely succeed, or should be able to recover from a faulty install. A failure in the procedure should be be make sure that the device can revert back to its original stable version. This is also called atomic update property.

\subsubsection{Code signing and Verification}
OTA updates should follow a certain set of rules, and cryptographic code signing should take place so as to make sure that the downloaded fimrware is not tampered with or corrupt in the first place.
\subsubsection{Code compatibility Checks}
Provisions to check whether the given firmware is built for the said system needs to be checked thoroughly before installation. MCUs with similar processor families may have similar toolchains to build the firmwares, however due to the fact that each processor has specific properties, it can cause runtime errors. These need to be averted at any cost.
\subsubsection{Partial updates should be the norm}
Partial updates reduce the bandwidth consumption, on-device processing time and reduces the possibilities of overwriting the firmware image entirely. The parts which are needed to be updated at a given time are the only parts which get affected by the patch files.
\subsubsection{Use Secure Communication Channels}
OTA updates should be downloaded only over encrypted channels. This includes the TLS connection required during the connection between cloud and edge devices and the security between gateway and fleet of edge devices.

\section{Decentralized Network and uses}
\subsection{Blockchain and its Features}
Blockchain technology is an ingenious invention based on the concepts of decentralised computing, making use of cryptographically secure methods to generate immutable ledger systems. The basic idealogy that blockchain follows is: 
\subsubsection{Immutability}
The data being stored in the blockchain is immutable, that is, the data can not be editted once the it is stored and the transaction is accepted.
\subsubsection{Decentralized Eco-System}
The network doesn't rely on a single central server to serve data. Instead, a number of interconnected nodes which are part of the same blockchain tend to share the ledger, or the the data and make sure that data is stored in all the nodes. In case a node is acting wrong or contains wrong data, the other nodes can easily rectify this error and update the node.\\

When working with blockchain-based decentralised application development, we usually select a blockchain solution which is able to provide autonomous conditional runs, that is, it should be able to run basic logic without the requirement of a human benefactor. For this very reason, the paper will extend the scope of blockchains towards Ethereum Blockchain System to allow the use of Smart Contracts.  Ethereum supports the use of a Virtual Machine (EVM) which can run user-defined contract codes, and can be used to build Dapps or Decentralized Applications.

\subsection{Problems with Blockchain}
Blockchains are a relatively new technology, and while they can solve a bevy of problems, they also present certain new problems during development of a solution. Ethereum allows the use of multiple types of networks. The current MainNet or Main Network allows users to create and deploy contracts, however, the costs are expensive due to the fact that the price of gas (Ethereum's term for computation power) varies with change in market trends. This problem can be solved with the use of personal Ethereum chains. However, this will again increase the cost of maintenance as the user will have to deploy multiple servers which will act as nodes. Another problem arising from this is that if one user controls all the nodes of a blockchain, its pseudo-decentralized and is still vulnerable, unlike the MainNet.

\section{Architecture of the Blockchain Based OTA Service}
Taking the previous sections as the basis for the problem, it can be generalized that OTA services and server maintenance go hand-in-hand, but in real-time scenarios it is usually managed by two different entities. This creates a co-dependency where one device needs to trust the other network that the files being served are not malicious, and the server itself is completely secure. This false trust is what acts as a loophole.
The proposed architecture employs a blockchain based application, which can hold the details to a firmware in transactions. In simple words, the blockchain will store the firmware details of the latest version of the product, and any IoT device, which will check for updates can easily get the details directly by making a getter request to it through the smart contract. The getter details are completely open-ended, and any device can obtain it without the requirement of a specific key.

As Ethereum cannot hold firmware files and it is very expensive to actually hold such big files in a transaction, the firmware files will be uploaded to Inter-Planetary File System, which is a decentralized file system space. When the systems developer of the IoT product uploads the firmware to IPFS, a unique key is presented to the developer which can then be appended to the blockchain via a transaction. This unique key can be used to get files from the IPFS network directly via HTTP or web-socket interfaces. IPFS acts as a database as much as a file system due to the fact that all the files are linked to hashes and are directly stored to hash tables.
\subsection{Security Concerns}
One major problem regarding security is that the developer should be the only person or part of a group of people who are allowed to upload firmware files. If a malicious user is able to upload files to the blockchain, then the point of using a smart contract will be completely void. To solve this problem, the contract employs a multisig wallet. MultiSig wallets are a set of addresses generated from a single root, and are very similar to HD wallets. All these wallet addresses are allowed to act as co-owners to the contract, and are the only ones allowed to execute the functions defined in the contract such as uploading the firmware.

\begin{figure}[ht!]
  \includegraphics[width=\linewidth]{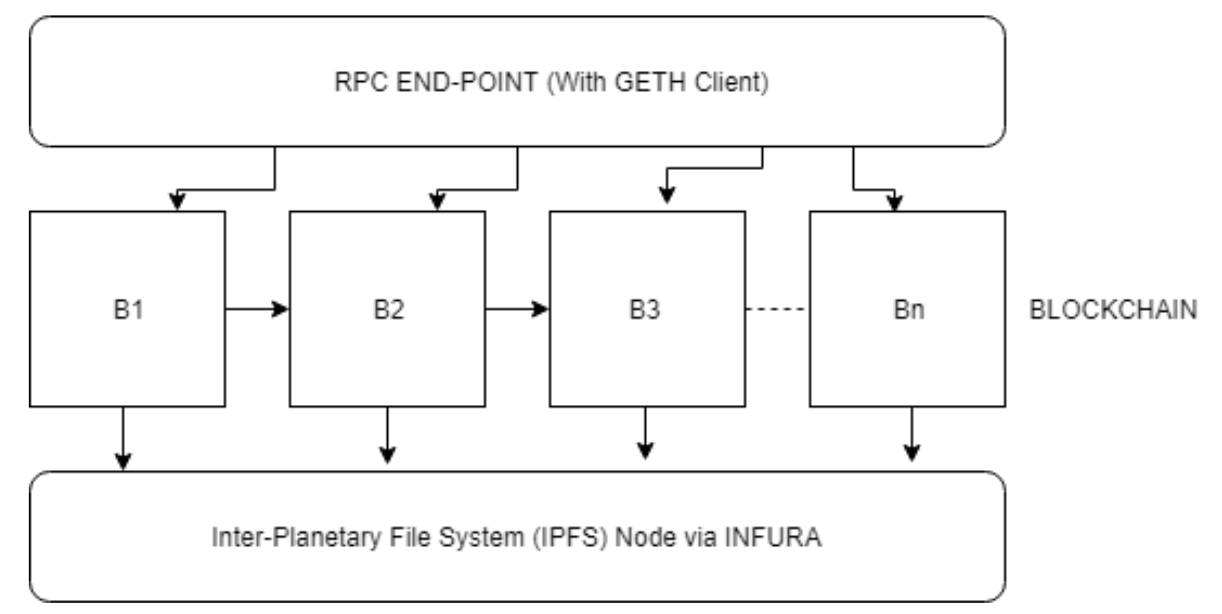}
  \caption{Blockchain Architecture of OTA Service}
  \label{fig:PIC2}
\end{figure}

\section{Testing of Blockchain based OTA Service}
From the proposed blockchain architecture, a smart contract can be modelled to run the required functions according to given constraints. The IPFS data and other metadata is usually saved in the form of u256-bit strings, whereas the data for incremental versions is saved in the form of unsigned integers (uint8 upto uint256).  The smart contract was simulated over ganache core which can host a single node test-rpc client for personal ethereum network. The basic call timing alongside delay in function calls are given below.

\begin{figure}[ht!]
  \includegraphics[width=\linewidth]{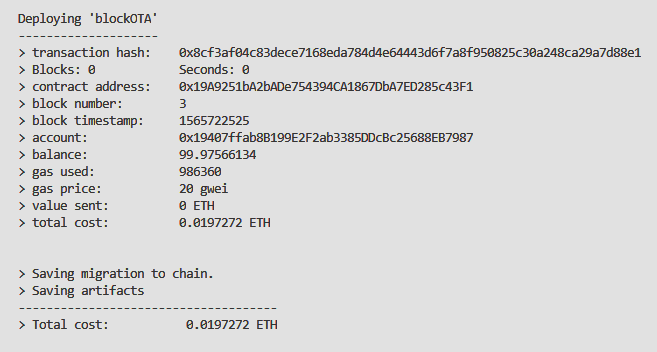}
  \caption{Deployment Statistics}
  \label{fig:PIC3}
\end{figure}

The delay test was simulated using chai-assertions and passing the test through truffle development framework.

\begin{figure}[ht!]
  \includegraphics[width=\linewidth]{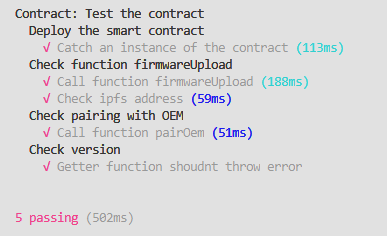}
  \caption{Time Delay over Personal Ethereum Network}
  \label{fig:PIC4}
\end{figure}

As the test denotes, the function calls for getter function is 91ms, which is pretty acceptable when working with personal networks. The Firmware upload function requires to load 5 arguments to save the firmware files, and denotes the exact delay required to call the function.\\

Another important aspect of the system's throughput will be defined by the download rate of the firmware files from an IPFS end-point. This was simulated by using 4 different geographic locations as server end-points, and the latency of the download is compared to HTTPS 2.0, which is the current norm. The same payload was used for the simulation of all the cases.

\begin{figure}[ht!]
  \includegraphics[width=\linewidth]{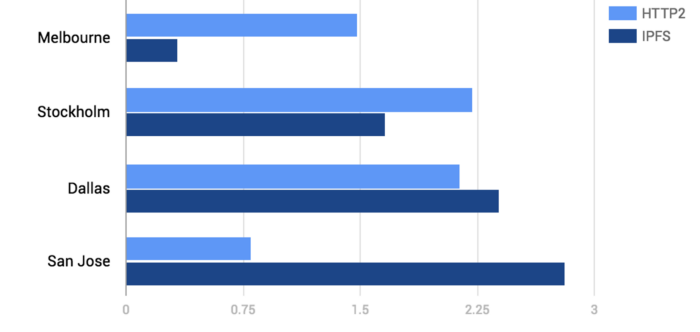}
  \caption{Comparison of latency between HTTPS and IPFS}
  \label{fig:PIC5}
\end{figure}

From the bar graph, it can be assumed that IPFS latency completely depends on the location of the call. The calls were made from San Francisco, CA and correlating it to the given data, we see that HTTP is faster for neighbouring cities. However, as the distance increases, the delay increases proportionally, and after a point, IPFS tends to provide higher throughput. But when compared this data on a more simple level, we can easily assume that the latency is close to HTTP2, hence proving to be quite viable when used for storing firmwares.

\section{Hardware Implementation}

\begin{figure}[ht!]
  \includegraphics[width=\linewidth]{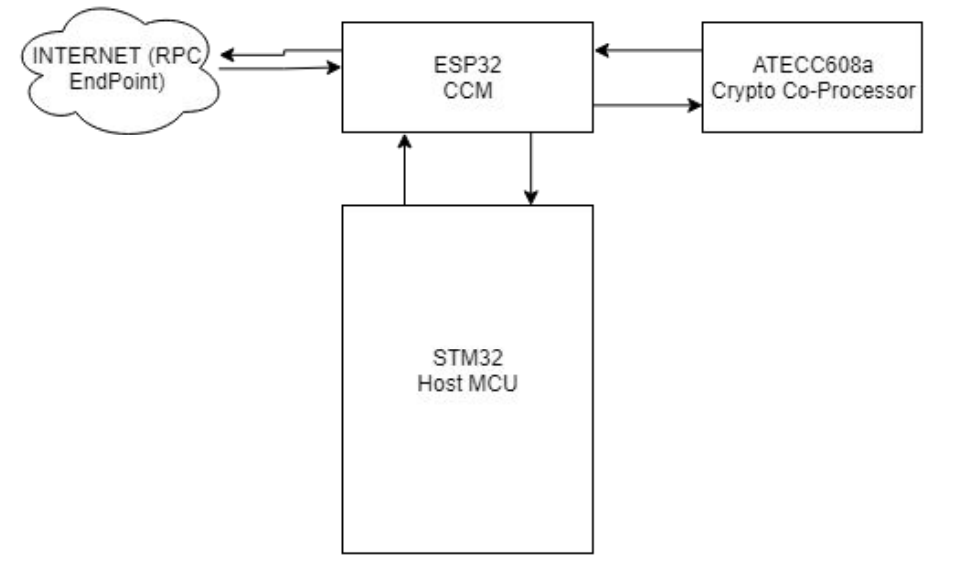}
  \caption{Hardware Design Block Diagram}
  \label{fig:PIC6}
\end{figure}

IoT devices are designed using the basic ideology that it should contain all the necessary network interfaces to connect to any IoT gateway. This includes RF communication links such as Bluetooth and XBee, or WiFi communication interfaces such as ESPressif's ESP MCUs. For the purpose of studying the system in real-time, the implementation will use the following components
\subsubsection{ESP8266 Microcontroller}
ESP8266 is a 32-bit MCU capable of using WiFi and communication protocols such as MQTT (Message Queuing Telemetry Transport) which enables the system to communicate with the blockchain. The MCU also has a 4MB SPIFFS storage, which will be used to store downloaded firmwares.
\subsubsection{STM32F103 Microcontroller}
STM32F103 Microcontroller, commonly known as the BluePill, will act as the host MCU for the system. The bluepill will emulate the active MCU system, on which the firmware will be regularly updated.
\subsubsection{ATECC608a Crypto Co-Processor}
ATECC608a is a crypto co-processor modules, which is available in DIP-8 packages. The co-processor will be used to maintain integrity of the system and will be used during the update and boot process to make sure that the firmware is not corrupted. The co-processor will communicate using I2C interface and doesn't cause bottlenecks.

\subsection{Requirement of Crypto Co-processor}
ATECC608a is the selected cryptographic co-processor that is used in the hardware implementation. It supports elliptic curve algorithm, which can be used to enhance the security of IoT devices with the help of authentication services. The component has a protected storage for storing up to 16 keys and certificates and also supports Diffie-Hellman key exchange. In the given implementation constraints, the device can function as following 

\subsubsection{Cloud Authenticator}
This is used to authenticate the security of the cloud or the gateway the IoT device will communicate with. The certificates stored in the component will be used for handshakes and ESP8266 by-itself can sustain TLS connections over MQTT
\subsubsection{Secure Boot Checker}
Randomly check boot partitions while boot up process. When a small sector of flash is taken and hashed using the co-processor, the hash value should match with the one present in the protected memory region of the co-processor.
\subsubsection{Offline Transaction Signing Component}
Protected private keys, which are part of the Ethereum wallet can be stored securely in the co-processor and can be used to sign any transactions in case the IoT device intends to send data to the blockchain network.
\subsubsection{Hash Authenticator}
The easiest way to check whether two quantities of data are the same is to pass both of them through a non-reversible hash function and then comparing the strings directly. If the hash values match, then the input data 1 was the same as input data 2. 
This can be used to securely check the flash sectors of an embedded system during boot or can also be used to verify firmware before installation by hashing it directly and checking it against the hash value taken from the blockchain.\\

Hence for a hash authenticator, if
\begin{equation*}
  H(a) = H(b)
\end{equation*}
then it can be verified that
\begin{equation*}
  a = b
\end{equation*}
The co-processor accelerates the computational time required to generate hash values. In the above equation, H(x) denotes a hash function, with an input x. Usually SHA256 or its derivative is used. For two given inputs a and b, if the above equation is valid then we can conclude the a and b are the same.

\subsection{Practical Design On CAD Software}
Practical design rules apply very well to the design of the system. The system requires minimal connections to allow the OTA process. However, length matching of lines is highly important when working with the serial and I2C lines, so as to not cause any bottlenecks for a given baud rate.

\begin{figure}[ht!]
  \includegraphics[width=\linewidth]{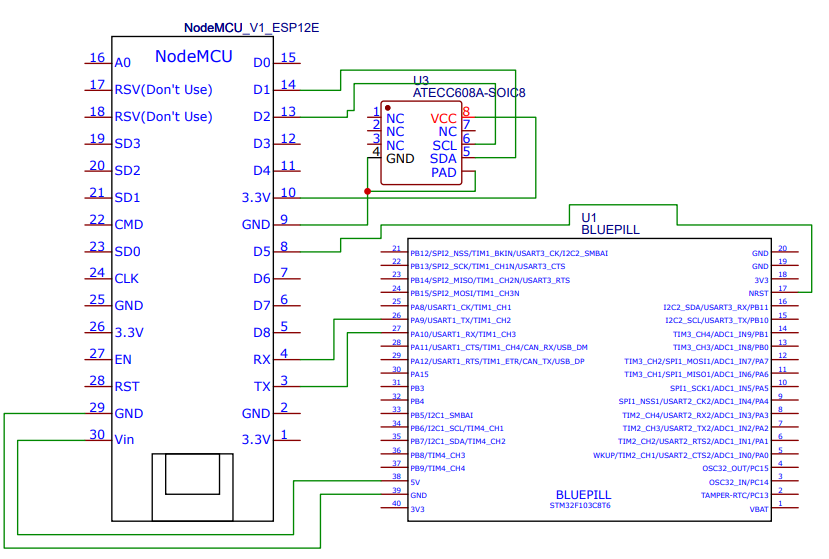}
  \caption{Basic Schematic of Test Hardware}
  \label{fig:PIC8}
\end{figure}

\subsection{Update Procedure from Hardware side}
Once the STM32 device is initialized in boot-loader mode by pulling the BOOT pin to ground, the USART\_Rx1 line starts sensing for the start command, that is "0x7F". Once it receives the command, the ESP8266 checks whether the system is write protected or not. If the system is not protected, it continues the process and retrieves device information, to check for code provenance. On passing the test, the esp8266 start sending data frames containing the firmware, and rewrites the flash memory from a given address location to an end location address. Being able to control the start and end address allows the device to partially update without erasing other memory locations.

\begin{figure}[ht!]
  \includegraphics[width=\linewidth]{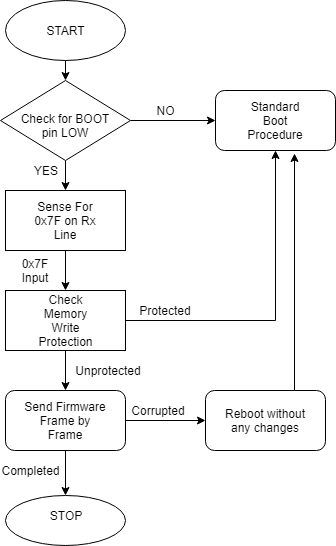}
  \caption{Hardware update procedure block diagram}
  \label{fig:PIC7}
\end{figure}

\section{Conclusion}
During the final test results, the device performance was tested over Ropsten Test Network, which gives us an accurate estimation of how much time will the entire process take.

\begin{table}[h!]
  \begin{center}
    \caption{Final Test Results}
    \label{tab:table1}
    \begin{tabular}{l|r} 
      \textbf{Name} & \textbf{Time Delay}\\
      \hline
      Contract Deploy & 103 Seconds\\
      Transaction Call & 4 Seconds \\
      Transaction Confirm & 63 Seconds \\
      Getter Call & 7 Seconds\\
    \end{tabular}
  \end{center}
\end{table}

The results in TABLE 1 shows that the system will be quite usable in a real-time scenario. With the problems and possible solutions discussed in the above sections, it can be assumed that the proposed system is secure, functional and solves many problems posed by the original OTA service methods.

\section{Acknowledgements}
The Author would like to thank Dr. S. Dhanalakshmi, for guiding the research project, and providing an opportunity and space to implement novel ideas. The author extends his gratitude towards ST Microelectronics and Microchip for providing ample information in the form of datasheets and technical documentations.


%

\ifCLASSOPTIONcaptionsoff
  \newpage
\fi



%

%

\begin{IEEEbiographynophoto}{Projjal Gupta}
is a final year student with the Department of Electronics and Communication department. He has worked on multiple hardware devices and development boards, and has research interests in the field of Internet of Things and Security, VLSI, Embedded Systems, RTOS and decentralized computing.
\end{IEEEbiographynophoto}





\end{document}